\documentstyle[prl,aps,multicol,epsf]{revtex}
\input{psfig}

\begin{document}

\bibliographystyle{simpl1}

\title{Andreev Scattering and the Kondo Effect}

\author
{Aashish A. Clerk$^{1}$, Vinay Ambegaokar$^{1}$, and Selman Hershfield$^{2}$} 

\address
{$^{1}$Laboratory of Atomic and Solid State Physics,
Cornell University, Ithaca NY 14853, USA,
 $^{2}$Department of Physics, University of Florida, Gainesville FL 32611, USA}
\maketitle

\centerline{(July 14, 1999)}
 
\begin{abstract}

The properties of an infinite-$U$ Anderson impurity coupled
to both normal and superconducting metals is studied using a generalization 
of the noncrossing approximation which incorporates multiple Andreev 
reflection.  Both the cases of a quantum dot and a quantum 
point contact containing an impurity are considered.  We find that the 
magnitude of the Kondo resonance is altered, and that structure 
develops at energies corresponding to the superconducting gap.  This 
leads to observable changes in the zero-bias conductance.  We also
find that magnetic and non-magnetic Kondo effects respond in opposite
manners to Andreev reflection.  
\end{abstract}
\draft


\begin{multicols}{2}

\section{Introduction}

Though the Kondo effect is far from being a new problem in condensed matter 
physics, the question of what occurs when one places a Kondo impurity on the
normal side of a normal-superconducting (NS) interface has only recently 
received attention \cite{golub,fazio.eom,kang,fazio.mft}.
The NS case is significantly different from the 
more studied problem of a magnetic impurity in a bulk superconductor, in 
which a lack of low-lying excitations leads to a suppression of the Kondo 
effect.  In the present problem, we have both the existence of low lying 
excitations (provided by the normal metal) and anomalous pair correlations 
(a result of the proximity effect).  The question of what influence this 
combination will have on the dynamically-generated Kondo effect is the subject
of this work.  In particular, how will the sharp Kondo resonance at the Fermi 
energy be modified?  

It should be noted that the immediate motivation for studying this problem 
has come from experiment.  Recent results obtained from semiconductor quantum 
dots have shown clear signs of the Kondo effect \cite{goldhab,otherdot}, 
leading to the question 
of what one would expect if a quantum dot were now coupled to both normal and 
superconducting leads.  In another class of systems, zero-bias conductance 
anomalies seen in metallic quantum point contacts have been attributed to 
scattering 
off effective non-magnetic, two channel Kondo impurities \cite{tls,titanium}.  
Such point contacts can be constructed between normal and superconducting 
metals \cite{shashi}, providing further relevance for the current problem.

\section{The Models}

In this work, we study three different systems:  a quantum dot coupled to both
normal and superconducting leads (NS-QDOT), a quantum point contact between
normal and superconducting metals which contains a magnetic impurity
(M-QPC), and a normal-superconducting point contact which contains a 
non-magnetic, two-channel Kondo impurity (NM-QPC).  We outline in this section 
the models used for each of these devices.

For the first system, the NS-QDOT, we treat the quantum dot as an 
infinite-$U$ Anderson impurity coupled to both normal and 
superconducting leads.  The infinite-$U$ Anderson model has been used
successfully in the absence of superconductivity to describe the Kondo 
effect in quantum dots; our model represents a natural generalization to the
NS case.  Using a slave-boson representation, we have:  

\begin{eqnarray}
H_{dot}= H_0+\varepsilon _d \sum _{\sigma } f^{\dag}_{\sigma }
f^{\phantom{\dag}}_{\sigma } + W\sum _{\alpha ,k,\sigma} 
(c^{\dag}_{\alpha ,k\sigma}b^{\dag } f^{\phantom{\dag}}_{\sigma} + h.c.),
\label{HDOT}
\end{eqnarray}
\begin{eqnarray}
H_0 = \sum _{\alpha,k,\sigma} \varepsilon _k 
c_{\alpha,k\sigma}^{\dag}c^{\phantom{\dag}}_{\alpha,k\sigma} + 
\sum _{k} (\Delta c^{\dag}_{S,k\uparrow} c^{\dag}_{S,-k\downarrow} + h.c. ).
\label{HO}
\end{eqnarray}
{\noindent}The $c^{\dag}_{\alpha ,k\sigma}$ operators here
create band electrons, with 
$\sigma$ denoting spin and $\alpha =N,S$ labelling the two leads.  $\Delta$
represents the pair potential in the superconducting lead.  The
Anderson impurity has bare energy $\varepsilon _d$, and is represented
in the usual manner using auxiliary fermion ($f$) and boson ($b$) operators;
the $U = \infty$ constraint of single occupancy takes the form 
$\sum _{\sigma } f^{\dag}_{\sigma }f^{\phantom{\dag}}_{\sigma } + b^{\dag}b
= 1$. 
 
For the NS point contact with magnetic impurity system (M-QPC), 
we use a model consisting of a rectangular wire of transverse 
dimensions $w$, extended in the $z$ direction and having a step-function 
pair potential 
$\Delta(z)=\Theta(z)\Delta$.  
The magnetic impurity is modeled as an infinite-$U$ 
Anderson impurity sitting at a point {\bf $a$} on the NS interface:
\begin{eqnarray}
H_{m} = H_{Q0} + \varepsilon _d \sum _{\sigma } f^{\dag}_{\sigma }
f^{\phantom{\dag}}_{\sigma } + W \sum _{\sigma} 
(\Psi^{\dag}_{\sigma}({\bf a})b^{\dag } f^{\phantom{\dag}}_{\sigma} + h.c.)
\label{HQPC}
\end{eqnarray}
where $\Psi^{\dag}_{\sigma}({\bf x})$ creates a band electron at position 
${\bf x}$ and
\begin{eqnarray}
H_{Q0} = \int d{\bf x} \biggl( \sum _{\sigma}\Psi^{\dag}_{\sigma}({\bf x})
(\frac{-\hbar^2 \nabla^2}{2m} - E_F)
\Psi^{\phantom{\dag}}_{\sigma}({\bf x}) + 
\label{HOQPC}
\\
\Delta(z) 
\Psi^{\dag}_{\uparrow}({\bf x})
\Psi^{\dag}_{\downarrow}({\bf x}) + hc. \biggr). 
\nonumber
\end{eqnarray}

For the final system of an NS point contact containing a non-magnetic, 
2-channel impurity (NM-QPC), we use a model similar to that for the M-QPC
system.  Now, however, the impurity represents a two-level system (TLS) 
off which electrons may scatter causing level-flips of the TLS 
\cite{pseudoref}.  In this model, 
the impurity does not carry ordinary spin, but rather has a pseudo-spin index 
$\tau$ which can take on one of two values, corresponding to the two
states of the TLS.  The interaction here 
between impurity and conduction electrons is also non-magnetic, 
meaning that electron spin
must be conserved.  This is accomplished by having the slave-bosons carry a 
spin index.  We have:

\begin{eqnarray}
H_{nm} = H_{Q0} + \varepsilon _d \sum _{\sigma } f^{\dag}_{\sigma }
f^{\phantom{\dag}}_{\sigma } + W \sum _{\sigma,\tau} 
(\Psi^{\dag}_{\sigma,\tau}({\bf a})b^{\dag }_{\sigma} 
f^{\phantom{\dag}}_{\tau} + h.c.)
\label{HNMQPC}
\end{eqnarray}
{\noindent} where the conduction 
electron Hamiltonian $H_{Q0}$ is as in (\ref{HOQPC}),
with the modification that the conduction electron operators also carry 
the pseudo spin index $\tau$. We take the pairing in (\ref{HOQPC})
to be diagonal in this index,
which is compatible with the usual interpretation of the the pseudo-spin as a 
parity index.

It is important to note that there are significant differences between
the three systems we study, as can be seen from their respective Hamiltonians.
In the NS-DOT system, the N and S leads are {\it only} in contact through
the impurity; all transport through the system will involve it.  
In the point contact systems, we have the
opposite situation--  the N and the S metals are in 
perfect contact, with the impurity acting only as an additional source of 
scattering at the interface.  

The difference between the two point contact systems is
also worth emphasizing.  In the M-QPC system, our impurity is magnetic
and thus conduction electron spin is not conserved-- real spin flips
can occur.  In the NM-QPC system, 
conduction electron spin {\it is} conserved; only the auxiliary
pseudo-spin index $\tau$ can undergo spin flips.  This difference will
prove to be crucial, due to the sensitivity to spin ordering of the singlet
pairing in the superconductor. 

\section{Extension of the Non-Crossing Approximation}

For all three of our systems, 
we calculate the impurity spectral function (also called the
impurity density of states) using an 
extension of the self-consistent non-crossing approximation (NCA) 
\cite{ncarefs}.  The NCA
amounts to an infinite re-summation of perturbation theory,  and has been 
shown to be quantitatively reliable down to temperatures well 
below $T_K$ \cite{ncarefs}.  We modify the NCA to now include 
multiple-Andreev reflection processes.  The resulting 
NS-NCA $f$-fermion and slave boson propagators are given in Fig. 1.
The new graphs in our approximation are those which contain anomalous
propagators; they do not appear in the usual NCA. 

Like the original NCA, the NS-NCA is a conserving approximation as it 
may be derived by differentiating a generating functional (i.e.. it is 
$\Phi$-derivable) \cite{phideriv}.  Moreover, if $N$ is 
the degeneracy of the Anderson 
impurity, the NS-NCA includes {\it all} graphs to order
$1/N$, including new Andreev reflection graphs which first 
appear at this order.  
As the success of the normal NCA is attributed to the fact
that it too is exact to order $1/N$ (in the absence of superconductivity), 
the NS-NCA 
employed here can be viewed as a natural extension to NS Kondo systems
\cite{largeN}.   

\begin{figure}
\centerline{\psfig{figure=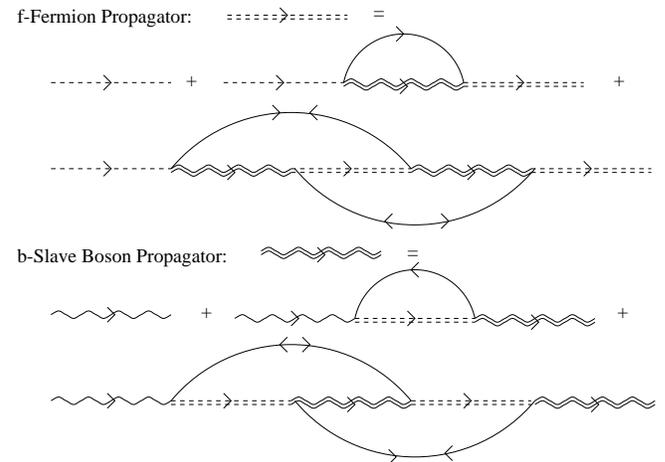,width=8.5cm}}
\narrowtext
\caption{
Diagrammatic representation of the NS-NCA.  Dashed lines are f-fermions,
wavy lines are slave bosons, solid lines are lead electrons. Double lines
indicate a fully dressed propagator.  Anomalous propagators indicate 
Andreev reflection.
}
\label{nsnca}
\end{figure}

It should be noted that previous studies of NS Kondo systems have either
completely neglected changes in impurity dynamics due to superconductivity 
\cite{golub}, or have relied on the equations of motion approach which 
is formally only valid at temperatures above the Kondo temperature 
\cite{fazio.eom}.  In analogy to the usual NCA, we expect our approach 
based on the NS-NCA to be quantitatively valid for temperatures well 
below $T_K$.

The Dyson equations pictured diagrammatically in Fig. 1 lead to a set of
coupled integral equations for the $f$-fermion and slave boson propagators. 
Letting $F(\omega ) =(\omega -\varepsilon _d - 
\Sigma (\omega ))^{-1}$ and $B(\omega ) 
=(\omega - \Pi(\omega ))^{-1}$ represent the $f$-fermion and slave boson 
retarded propagators respectively, the equations read:

\begin{eqnarray}
\Sigma(\omega) = \frac{M\Gamma}{\pi} \int d\varepsilon
\biggl( \rho (\epsilon) B(\omega - \varepsilon) f(-\varepsilon) \mp
\label{NSF}
\\
\frac{\Gamma}{\pi} \int d\varepsilon' 
\alpha(\varepsilon) \alpha(\varepsilon') 
B(\omega + \varepsilon) F(w + \varepsilon + \varepsilon')
B(\omega + \varepsilon') \biggr)
\nonumber
\end{eqnarray}

\begin{eqnarray}
\Pi(\omega) = \frac{2\Gamma}{\pi} \int d\varepsilon
\biggl( \rho (\epsilon) F(\omega + \varepsilon) f(\varepsilon) \pm
\label{NSB}
\\
\frac{\Gamma}{\pi} \int d\varepsilon' 
\alpha(\varepsilon) \alpha(\varepsilon') 
F(\omega + \varepsilon) B(w + \varepsilon + \varepsilon')
F(\omega + \varepsilon') \biggr).
\nonumber
\end{eqnarray}

In these equations, $M$ is the channel degeneracy of the impurity--
it equals one for both the NS-QDOT and the M-QPC systems, and equals two
for the NM-QPC system (where the two conserved values of electronic spin
play the role of the two channels).  The upper (lower) sign corresponds
to the $M=1$ ($M=2$) systems.
$\rho(\varepsilon)$ is the electronic density of 
states, $\Gamma=\pi W^2 \rho(0)$ the 
bare tunneling rate, $f$ the Fermi distribution function,  
and $\alpha(\varepsilon)$ is an
effective electron-hole coherence parameter defined by:
\begin{eqnarray}
\alpha(\omega) = \sum _{n} u^{*}_{n}({\bf a}) 
v^{\phantom{*}}_{n}({\bf a})
\delta (|\omega| - \varepsilon _{n}) f(\omega)
\label{alph}
\end{eqnarray}
where $u_n$ and $v_n$ are the usual BCS coherence factors. 

\begin{figure}
\centerline{\psfig{figure=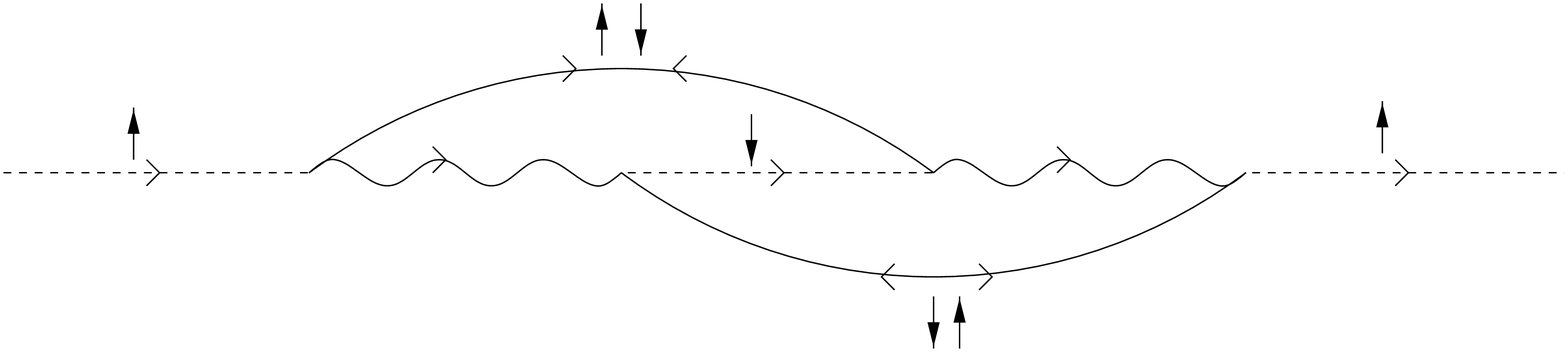,width=8.5cm}}
\centerline{\psfig{figure=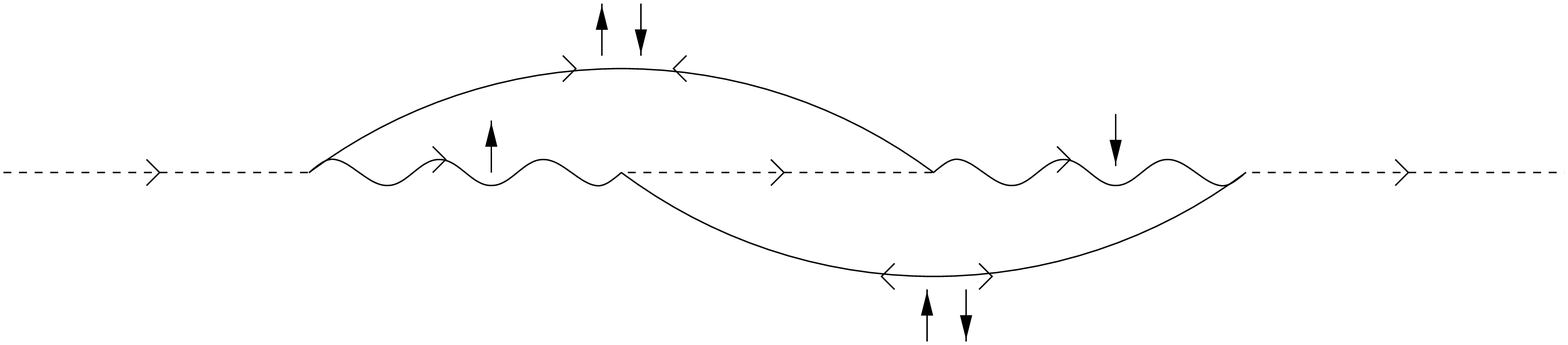,width=8.5cm}}
\narrowtext
\caption{
Diagrammatic argument for the different sign of the Andreev reflection
graphs in equations (\ref{NSF}).  The first line
shows a typical double Andreev reflection event for a magnetic
impurity.  The impurity carries a spin index in this case (i.e. the dashed
propagator), and we see that both Andreev reflections are related by
time reversal symmetry.  The second line shows the situation for a non-magnetic
impurity.  Here, the impurity itself does not carry a spin, but the slave-boson
(wavy line) does in order to conserve spin.  The two Andreev reflections in
this case are {\it not} time reversed partners; the second is off
by a spin rotation.
}
\label{grapharg}
\end{figure}

It is worth commenting at this point on why the new Andreev reflection terms in
equations (\ref{NSF}) and (\ref{NSB}) enter with a different sign for the 
NM-QPC system compared to the NS-QDOT and M-QPC systems.  The crucial
difference has nothing to do with the channel symmetry of the impurity (i.e. 2
versus. 1), but rather lies in the fact that the impurity-band interaction in the
NM-QPC is non-magnetic, whereas it is magnetic in the NS-QDOT and M-QPC 
systems.  The fact that conduction electron spin must be conserved in
the non-magnetic case 
has a direct consequence on the sign of the Andreev-reflection 
graph appearing in the $f$-fermion self energy-- 
the two Andreev reflections are not time-reversed partners, but are in fact 
off by a single spin rotation, leading to the additional
factor of $(-1)$ (recall that the BCS anomalous propagator changes sign
under a spin rotation).  This argument
is displayed graphically in Fig. \ref{grapharg}.  A similar argument can be
made to explain the sign difference occurring in the slave-boson graph.  

We thus see that the new Andreev
reflection processes which contribute to the dynamics of the impurity are
sensitive to whether or not the impurity is magnetic; this is a direct
consequence of the spin-sensitivity of the singlet pairing in the 
superconductor.  This difference will have a 
significant consequence on how the Kondo 
effect is modified, as will be discussed in what follows.      

\section{Results}
In all three of the systems considered, we choose model parameters 
corresponding to having the Anderson impurity in the Kondo
regime, and use a Gaussian with half-width D for the normal-state
density of states.  Our choices of $\varepsilon _d = -0.67D$, $\Gamma=0.15D$ 
yield  
$T_K=0.0005D$ for a one-channel impurity (i.e. the NS-QDOT and M-QPC systems),
and $T_K=0.0001D$ for a two-channel impurity (i.e. the NM-QPC system).  
We have numerically solved the 
NS-NCA equations in equilibrium for various 
temperatures and values of the superconducting gap.  Within the NCA, the 
impurity spectral function $A_D(w)$ can be directly related to the $f$-fermion
and slave boson spectral functions:

\begin{eqnarray}
A_{d\sigma}(\omega) = \int {d\varepsilon} [{\rm e}^{-\beta\varepsilon} + 
{\rm e}^{-\beta(\varepsilon - \omega)}] 
A_{f\sigma}(\varepsilon)A_{b}(\varepsilon - \omega)
\label{AD}
\end{eqnarray}
where the auxiliary particle spectral functions are 
defined by $A_{f} = - \frac{1}{\pi} {\rm Im} \, F$, 
$A_{b} = - \frac{1}{\pi} {\rm Im} \, B$. 
{\noindent} Note that the equality in equation (\ref{AD}) reflects a neglect
of vertex corrections which is consistent with the large-$N$ nature of the
NCA.  

\subsection{NS Quantum Dot}
\begin{figure}
\centerline{\psfig{figure=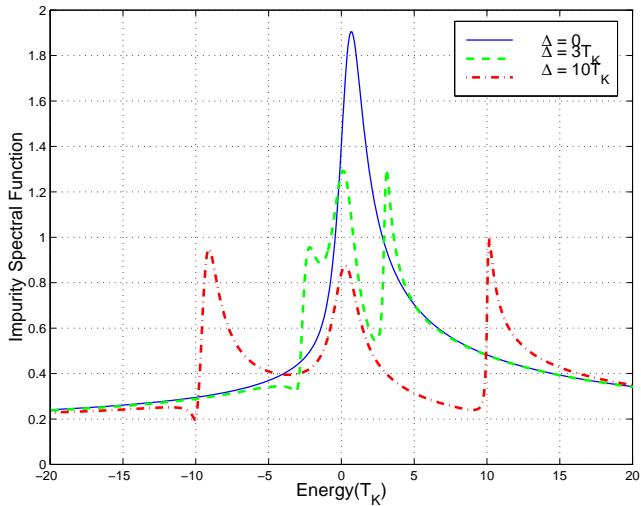,width=8.5cm}}
\narrowtext
\caption{
NS QDOT spectral function $A_{d\sigma}(\omega )$ for various values
of $\Delta $ at $T=0.5T_K$
}
\label{qdotres}
\end{figure}

In Fig. \ref{qdotres}, we plot calculated spectral functions for the 
NS-QDOT system for several values of the superconducting gap $\Delta$.  
In the case of no
superconductivity ($\Delta=0$), we see as expected the emergence of the sharp
Kondo resonance at the Fermi energy.  As the superconductivity is gradually
turned on, several interesting modifications of this resonance are observed.

The Kondo resonance does not vanish completely as $\Delta$ increases,
 though the spectral weight associated
with the resonance clearly decreases, indicating a partial suppression of the
Kondo effect.  This is to be expected when one recalls 
that the ground state of a magnetic, 
one-channel Kondo impurity is 
one in which the local impurity screens its spin with 
low-energy conduction electrons.  Turning on the superconductivity leads to 
a gap in the superconducting lead, and thus there are fewer low energy 
excitations available to the impurity to use in screening.  It thus
becomes more difficult to form the spin-screened Kondo ground state, resulting
in a diminishing of the Kondo effect.  A similar result for the NS QDOT
is found using slave-boson mean-field theory \cite{fazio.mft}, where the 
effective Kondo temperature is found to fall as $\Delta$ is increased above 
$T_K$.

Also of interest are the sub-peaks which develop in the impurity
spectral function at roughly $\pm \Delta$.  These peaks
are a result of the Kondo effect, and do not appear at higher 
temperatures.  Recalling that the the impurity spectral function at $0$ 
temperature can be viewed as a local particle addition/removal spectrum, 
these peaks indicate that one can create a superconducting-like excited 
state with excitation energy $\Delta$ by adding a particle to the 
quantum dot.  This indicates that {\it correlations are developing between
the superconductor and the impurity}, implying that the
superconducting electrons do indeed participate in the Kondo 
effect, even for $\Delta >> T_K$.  This can be substantiated by using
the large-$N$ variational approach \cite{GUNN} 
to calculate the approximate ground state of the NS-QDOT.  One finds that
there are superconducting quasiparticles present in the interacting ground
state, with a weight that scales as $T_K / \Delta$ for large $\Delta$.   

\begin{figure}
\centerline{\psfig{figure=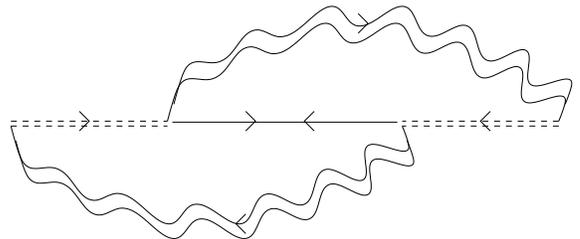,width=7.5cm}}
\narrowtext
\caption{
Diagram used to calculate anomalous impurity propagator within the 
NCA.  Double-dashed lines represent fully dressed f-fermion, double
wavy lines represent dressed slave boson propagators; both are
calculated within the NS-NCA.  The solid line represents an anomalous
lead electron propagator.
}
\label{anomgraph}
\end{figure}

As we find that a Kondo resonance persists in the NS-QDOT system, a natural
question to ask is whether there is any resulting enhancement of the zero-bias 
conductance of the device, as is seen in the absence of superconductivity.  
The equilibrium spectral functions obtained for the NS-QDOT can be used
to calculate the zero-bias Andreev conductance of the device, using an
approximate formula derived in \cite{fazio.eom}:

\begin{eqnarray}
G_{NS} =  \frac{4e^2}{h}\Gamma 
\int d\omega {\rm Re} \Sigma^{R}_{12} {\rm Im} 
(D^{R}_{12} D^{A}_{11})   [-\frac{df}{d\omega}]
\label{QDOTCOND}
\end{eqnarray}
Here, $D^{R(A)}_{ij}$ labels a component of the retarded (advanced)
impurity green function in Nambu-space,
and $\Sigma^{R(A)}_{ij}$ is the corresponding self-energy.  We have 
computed (\ref{QDOTCOND}) within the NS-NCA, evaluating the anomalous
impurity green function from the diagram in Fig. \ref{anomgraph} 
without use of the standard ``elastic'' approximation \cite{bickers}.  

For $T_K < \Delta < \Gamma$ and temperatures as low as one-half $T_K$,
we find that the NS QDOT {\em does not} exhibit a Kondo-induced
enhancement of the zero-bias conductance, despite the presence of a Kondo
resonance.  This can be understood by the fact that in the infinite
$U$ limit, there is a suppression of pairing at the dot compared to the $U=0$
case.  For energies and voltages smaller than $\Delta$, the only processes to
contribute to the current of the NS QDOT are those in which two particles
tunnel coherently through the dot and enter the superconductor as a pair.  The
suppression of pairing at the dot thus suppresses the zero bias conductance, 
despite the fact that there still is a Kondo resonance in the impurity 
spectral function.  Note that the same conclusion is obtained using a correct 
equations-of-motion approach (see the erratum in Ref. \cite{fazio.eom}).

\subsection{NS Point Contact Systems}

\begin{figure}
\centerline{\psfig{figure=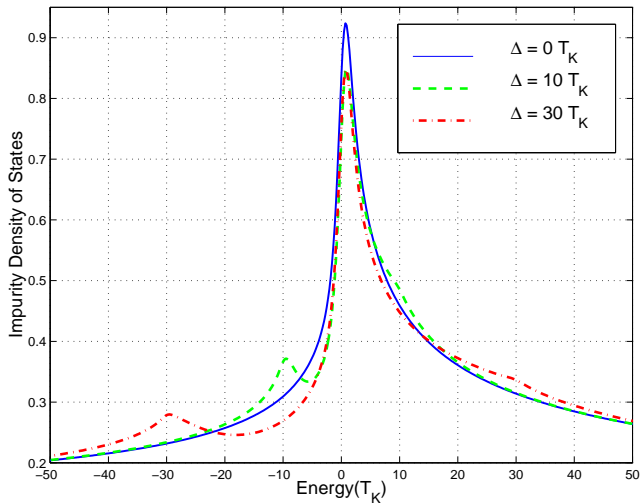,width=8.5cm}}
\narrowtext
\caption{
Impurity spectral function $A_{d\sigma}(\omega )$ for a non-magnetic
two channel Anderson impurity in an NS QPC for various values of $\Delta $ 
at $T=0.66T_K$.  Note that the introduction of superconductivity reduces the
height of the Kondo resonance.
}
\label{qpcres}
\end{figure}

\begin{figure}
\centerline{\psfig{figure=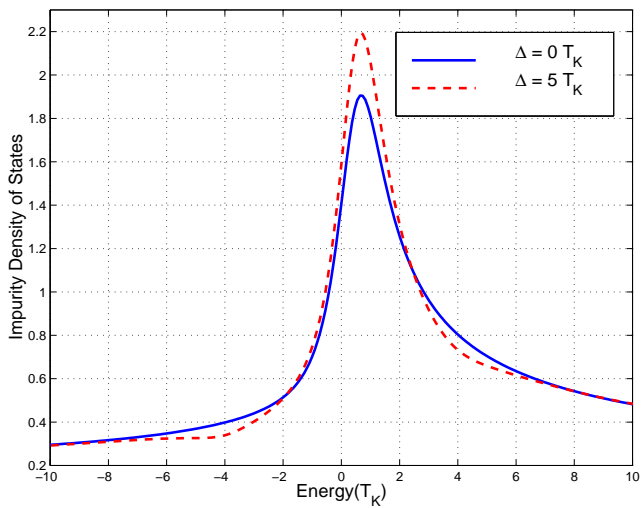,width=8.5cm}}
\narrowtext
\caption{
Impurity spectral function $A_{d\sigma}(\omega )$ for a magnetic
one channel Anderson impurity in an NS QPC at $T=0.50T_K$. Note here
that the introduction of superconductivity increases the height of the
Kondo resonance.}
\label{mqpcres}
\end{figure}

We proceed to discuss results for the NS point contact with magnetic
impurity (M-QPC) and NS point contact with non-magnetic impurity (NM-QPC)
systems.  
Shown in Fig. \ref{qpcres} is a plot of the impurity spectral function 
for the NM-QPC system for various values of $\Delta$ and at a temperature
below the bare Kondo temperature; Fig. \ref{mqpcres} 
shows a similar plot for the M-QPC system.

Several features are noteworthy in these plots.  First, we notice that in both
systems the introduction of superconductivity 
does not cause a significant change
in the amount of spectral weight in the Kondo resonance.  This indicates
that there is no major suppression of the Kondo effect here, in
contrast to the behaviour observed in the NS-QDOT.  This is to be 
expected-- for a clean, ballistic NS point contact, 
an energy gap does not form at the position of the impurity (which is 
at the interface) \cite{BDG}, and consequently there is no shortage 
of the low energy excitations needed to generate the Kondo effect.  
This differs from the NS QDOT, where the impurity is coupled directly to a 
bulk superconductor having a fully formed gap.


More interestingly, we find that in both point contact systems there
is a small modification of the height of the Kondo peak when $\Delta$ is
increased from $0$.  In the NM-QPC, we find that the Kondo peak is 
{\em reduced}, whereas the opposite is observed in the M-QPC-- here, the
Kondo peak is slightly {\em enhanced}.

This behaviour can be 
understood heuristically as follows.  First, note that by introducing 
superconductivity into the system, we introduce new processes which 
further couple the impurity to the conduction electrons 
(i.e. the multiple Andreev reflection processes which
correspond to the new graphs included in the NS-NCA).  These additional 
processes modify the effective coupling $\Gamma$ between the impurity 
and conduction electrons; an increase of this coupling will enhance
the Kondo effect, while a reduction will suppress it \cite{TKFORM}.  
In the present case, the sign of this modification depends on whether
or not the impurity is magnetic-- as discussed in Sec. III, 
the new Andreev-reflection graphs enter with opposite signs in these two 
cases due to the sensitivity of the Cooper pairing to spin ordering.
This leads directly to the 
opposite behaviour of the Kondo peak in the two cases.

The equilibrium impurity spectral functions can be used to derive the zero
bias conductance for the point contact systems.  
We compute the current 
using the non-equilibrium Keldysh technique as was done in 
\cite{wingreen,hershfield}, restricting ourselves 
to the regime where $V,T<< \Delta$.  
Using particle-hole symmetry, the zero bias conductance takes the form:
\begin{eqnarray}
G_{NS} = 2 G_{N} - \frac{4e^2}{h} \pi \Gamma 
\int d\omega A_d(\omega) [-\frac{df}{d\omega}(\omega)]
\label{QPCCOND}
\end{eqnarray}
where $G_{N}$ is the normal-state Sharvin conductance of the point 
contact in the absence of the impurity.  Details of this calculation
are presented in the appendix. Note that we
have neglected a term involving the anomalous impurity spectral function which
is $0$ for $T=0$ and is in general much smaller than the $A_d(\omega)$ term 
due to the strong suppression of on-site pairing.

The first term in (\ref{QPCCOND}) expresses the usual doubling of the 
conductance expected
for an ideal NS point contact, 
while the second term represents the correction
due to the impurity ($\delta G_{NS}$).  As in the N case, 
the formation of the Kondo resonance will lead to a small suppression of the
point contact conductance around zero voltage.  The correction term
due to the impurity in the absence of superconductivity ($\delta G_{N}$) 
has the same form as the second term in (\ref{QPCCOND}), but with an additional 
factor of $1/2$ \cite{hershfield}.  If Andreev reflection did not effect the 
dynamics of the impurity, $A_d(\omega)$ would be identical in both the NS 
and N cases.  We would thus expect the magnitude of the NS conductance 
anomaly to be {\it twice} that of the N anomaly.

\begin{figure}
\centerline{\psfig{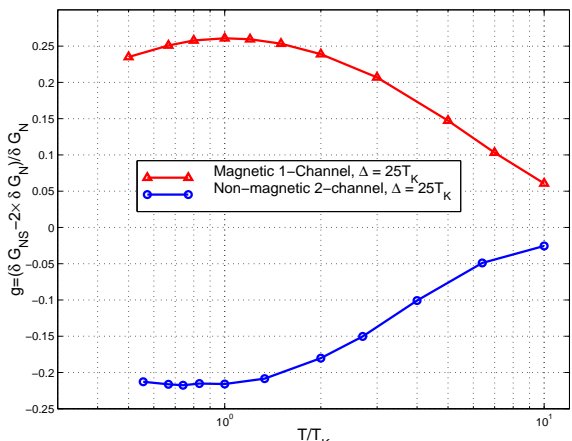}}
\narrowtext
\caption{
Parameter $g = (\delta G_{NS} - 2 \delta G_{N})/\delta G_{N}$ for 
the NS QPC as a function of temperature.  The non-zero value of $g$ 
is a consequence of changes in impurity dynamics due to Andreev reflection.
}
\label{qpcg}
\end{figure}

In Fig. \ref{qpcg}, we plot the parameter 
$g = (\delta G_{NS} - 2 \delta G_{N})/\delta G_{N}$ 
as a measure of the size of the NS QPC zero bias anomaly.  We see 
that as $T$ is lowered below $T_K$, $g$ becomes 
non-zero in both the cases of a magnetic and a non-magnetic 
impurity, indicating that Andreev reflection does indeed have an observable
impact on the Kondo effect in these systems.  Note also that
the sign of $g$ in both cases is in accord with our discussion of
the respective impurity spectral functions-- in the magnetic case, the NS
zero bias anomaly is {\it more} than two times the size of the anomaly in the
normal state, while in the non-magnetic case, the NS anomaly is {\it smaller}
than twice the normal state anomaly.  Again, magnetic and non-magnetic 
Kondo effects respond in opposite ways to Andreev reflection.
Significant too is the fact 
that the effect is large enough ($\pm 20\%$ at the lowest temperatures 
tested) that it should be experimentally accessible.  This behaviour 
could provide another test of the two-channel Kondo explanation of zero bias 
conductance anomalies seen in metal point contacts.   

\section{Conclusion}

We have studied the properties of three NS
Kondo systems using an extension of the NCA which incorporates
Andreev reflection.  We find the Kondo effect is indeed modified by
superconductivity in each system, leading to changes both in the impurity
spectral function and in the zero-bias conductance.  In the case of the 
NS-QDOT, superconductivity causes an overall decrease in the spectral 
weight of the Kondo resonance, and leads to the formation of new Kondo
sub-peaks at roughly $\pm \Delta$.  There is no resulting enhancement
of the zero-bias Andreev conductance due to the suppression of 
on-site pair
correlations in the large-$U$ limit.  In 
the point contact with impurity systems, we find that superconductivity
enhances the Kondo peak height in the case of a magnetic impurity, 
and suppresses it in the case of a non-magnetic impurity.  We attribute this
difference to the sensitivity of Cooper pairing to spin ordering.  

This work is supported in part by the NSF under grant DMR-9805613, and 
AC gratefully acknowledges financial support from the Olin Foundation.
AC and VA also thank the \O rsted Laboratory of the Niels Bohr Institute and NORDITA 
for their hospitality.   

\appendix
\section{Conductance of an NS Point Contact with Impurity}

We derive in this appendix Eq. (\ref{QPCCOND}) for the conductance of a
constriction between a normal metal and a superconductor which
contains an interacting impurity.  Our technique is
similar to that used in \cite{hershfield} to calculate the current of
a normal metal point contact with an impurity.  
Here, however, the ``trick'' of combining
currents calculated to the left and right of the constriction to
eliminate the impurity lesser function cannot be used-- we do not wish to
deal with the complexities of calculating the supercurrent on the S side
of our constriction.  We instead focus solely on the current in the normal
region, and use an approximate electron-hole symmetry which is present at low
energies to eliminate the impurity lesser function.

We begin by writing the current in terms of the lesser Green function:

\begin{equation}
\label{generalI}
I = -2 i e \hbar 
       \int \frac {d\omega}{2\pi} \int d^2x_{\perp} 
        \left(\frac {\partial _z - \partial _{z'}}{2mi}\right)
        G^<({\bf x},{\bf x}';\omega )\Big| _{{\bf x}={\bf x}'}.
\end{equation}
where $G^<({\bf x},{\bf x}';\omega )$ is the Fourier transform in time of:
\begin{equation}
\label{Glesserdefn}
  G^<({\bf x},t;{\bf x}',0) = i \left\langle 
       \Psi ^{\dag}_{\uparrow}({\bf x}',0) 
       \Psi _{\uparrow}({\bf x},t) \right\rangle 
\end{equation}
Note that we use only the $11$ component of the matrix 
Nambu Green function in computing the current;
a factor of two is included in (\ref{generalI}) to account for spin.  Note
also that in what follows we neglect the pseudo-spin index $\tau$ which
appears in the case of a two-channel Kondo impurity; our discussion may
nevertheless be applied to this case by
simply averaging over this index.

The next step in the derivation 
is to express the current in terms of the quasi-particle
operators $\gamma_{n\sigma}$,$\gamma^\dag_{n\sigma}$ 
which appear when one makes a Bogolubov-de Gennes (B-dG) transformation
to diagonalize the non-impurity parts of the Hamiltonian
(i.e. $\Psi_{\uparrow}({\bf x}) = \sum_n 
   u_n({\bf x})\gamma_{n\uparrow} - v_n^*({\bf x})\gamma^\dag_{n\downarrow}
   $ etc.).
This leads to the
following identification:

\begin{equation}
\label{modeG}
 G^<({\bf x},{\bf x}';\omega ) =
	\left[ \hat\phi_{1,m} ({\bf x})\right]_i  
	\left[ \hat G^<_{mn}(\omega) \right]_{ij}
	\left[ \hat\phi^*_{1,n} ({\bf x}')\right]_j
\end{equation} 
with:
\begin{eqnarray}
 \hat\phi_{1,m} ({\bf x}) & = & \left(
    	\begin{array}{c}
		u_m({\bf x})      \\
		-v_m^*({\bf x}) 
	\end{array}
  \right)	\\
 \hat\phi_{2,m} ({\bf x}) & = & \left(
    	\begin{array}{c}
		v_m({\bf x})      \\
		u_m^*({\bf x}) 
	\end{array}
  \right)	
\end{eqnarray}
\begin{eqnarray}
 \hat G^<_{mn}(t) = i
  \left(
   \begin{array}{cc}
	\left\langle 
		\gamma^\dag_{n,\uparrow}(0) \gamma_{m,\uparrow}(t) 
	\right\rangle &
	\left\langle 
		\gamma_{n,\downarrow}(0) \gamma_{m,\uparrow}(t) 
	\right\rangle \\
	\left\langle 
		\gamma^\dag_{n,\uparrow}(0) \gamma^\dag_{m,\downarrow}(t) 
	\right\rangle &
	\left\langle 
		\gamma^\dag_{n,\downarrow}(0) \gamma_{m,\downarrow}(t) 
	\right\rangle 
   \end{array}
  \right)
\end{eqnarray}

In Eq. (\ref{modeG}), $i$ and $j$ are indices in particle-hole
space, whereas $m$ and $n$ label different quasiparticle modes; we use
the $\hat{ }$ symbol to denote structure in particle-hole space.  
All repeated indices are to be summed over.  Note that the B-dG
wavefunctions $u$ and $v$ have a particularly simple form for mode 
energies $\epsilon$
less than the gap $\Delta$ \cite{Beenakker}.  Letting $\bar{m}$ index
the transverse wavefunctions $\psi_{\bar{m}}({\bf x}_\perp)$, we have for 
incident-electron modes:

\begin{eqnarray}
\label{bdgemodes}
  \hat\phi_{1,m} = \hat\phi_{1,(\bar{m},\epsilon)} =
     \left( \begin{array}{c}
	\exp(i k_e (\epsilon) z) \\ 
	\alpha(\epsilon) \exp(i k_h (\epsilon) z) 
     \end{array} \right) 
 \psi_{\bar{m}}({\bf x}_\perp) 
\end{eqnarray}
where $k_e= \sqrt{\frac{2m}{\hbar^2} (E_F + \epsilon - E_{\bar{m}})}$,
 $k_h= \sqrt{\frac{2m}{\hbar^2}(E_F - \epsilon - E_{\bar{m}}) }$ and
 $\alpha = \exp(-i\arccos(\epsilon/\Delta))$ is the Andreev reflection phase.
$E_F$ denotes the Fermi energy, while $E_{\bar m}$ is the energy of the
transverse mode $\bar m$.  Incident-hole wavefunctions have a similar form:
\begin{eqnarray}
\label{bdghmodes}
  \hat\phi_{1,m} = \hat\phi_{1,(\bar{m},\epsilon)} =
     \left( \begin{array}{c}
	\alpha(\epsilon) \exp(-i k_e (\epsilon) z) \\ 
	\exp(-i k_h (\epsilon) z) 
     \end{array} \right) 
 \psi_{\bar m}({\bf x}_\perp) 
\end{eqnarray}

The problem now becomes one of computing the mode-space lesser functions
$\hat G^<_{mn}(\omega)$.  As the only non-trivial interaction in
our system is at the impurity itself (the quasiparticles only interact
through a hopping term), we may express the exact self-energy appearing
in the Dyson equation for $\hat G^<_{mn}(\omega)$ in terms of 
$\hat D(\omega)$, the Nambu matrix Green function describing the impurity.

The Dyson equation takes the form:
\begin{eqnarray}
\label{lessDyson}
   \hat G^<_{mn} & = & \delta_{mn} \hat G^{0,<}_{n} 
	+ \hat G^{0,R}_{m} \hat \Sigma^<_{mn} \hat G^{0,A}_{n} + \\
\nonumber
&&	\hat G^{0,R}_{m} \hat \Sigma^R_{mn} \hat G^{0,<}_{n}
	+ \hat G^{0,<}_{m} \hat \Sigma^A_{mn} \hat G^{0,A}_{n}
\end{eqnarray}
where the $0$ superscript indicates a Green function of the non-interacting
system, and the $R$ ($A$) superscript indicates a retarded (advanced) Green
function.  All functions in the above formula are to be evaluated at the
same frequency.  The self-energy matrix appearing in (\ref{lessDyson}) is
simply:

\begin{equation}
\label{lessSE}
   \left[\Sigma^R_{mn}(\omega)\right]_{ij} = \sum_{\beta,\beta'=1,2}
	\left[ \hat\phi^*_{\beta,m}({\bf a}) \right]_{i}
	D^R_{\beta \beta'}(\omega)
	\left[ \hat\phi_{\beta',n}({\bf a}) \right]_{j}
\end{equation}
with similar formulas for the advanced and lesser self-energies.  Recall
that ${\bf a}$ indicates the position of the impurity.  This simple
form for the self-energy reflects the fact that quasiparticles only interact
by hopping on and off the impurity site.

We proceed by substituting Eq. (\ref{lessDyson}) into Eq. (\ref{modeG}).
We evaluate the resulting expression in the limit where $z,z'$ both
tend to $-\infty$ (i.e. deep in the normal region).  
As the wave functions $u_m({\bf x})$ and $v_m({\bf x})$ have
a plane wave form, the terms required to calculate the current are
greatly simplified.  A further simplification arises when we restrict
attention only to those quasiparticle modes with energies less than 
$\Delta$; this is valid for voltages and temperatures which are much 
smaller than $\Delta$.  

After a fair amount of algebra, we obtain a relatively simple formula
for the current.  Ignoring terms involving the anomalous impurity
Green functions for the moment and writing $I = I_0 + \delta I$, we have:

\begin{equation}
\label{I0}
 I_0 = 
	\frac{2e}{h} \int d\omega \sum_{\bar{m},E_{\bar m}<E_F}
	  \left( f_e(\omega) - f_h(\omega) \right)  
\end{equation}
\begin{eqnarray}
\nonumber
 \delta I & = &
     -\frac{2e}{h} \int d\omega \left(\pi \Gamma(\omega)\right) 
     \biggl(\frac{1}{2\pi i} D^<_{11}(\omega) - A_{11}(\omega) f_h(\omega)\\
&& \hskip40pt 
+ \frac{1}{2\pi i} D^<_{22}(\omega) - A_{22}(\omega) f_h(\omega) \biggr)
\label{bigdelI}
\end{eqnarray}
In the above equations, $f_e(\omega)$ is the distribution function 
describing electrons, while $f_h(\omega) = 1 - f_e(-\omega)$ describes 
holes.  $A_{ij}(\omega) = \frac{1}{\pi}$Im$(D^R_{ij}(\omega))$ is an 
impurity spectral function.  $\Gamma(\omega) = \pi W^2 N(\omega)$ is an
energy-dependent tunneling rate, with $N(\omega)$ being the density of
states.  Of course, the energy integrals in both these equations should
be restricted to values less than $\Delta$ due to the approximations that
have been made; this will occur naturally for
sufficiently small temperatures and voltages.

$I_0$ in Eq. (\ref{I0}) is the point contact current in the absence of 
the impurity.  Taking $f_e(\omega)$ to be the shifted Fermi function 
$f(\omega - eV)$, we find that $G_0 = \frac{dI_0}{dV} = 
\frac{4e^2}{h}N_m$, where $N_m$ is
the number of transverse modes with energies below the Fermi energy.  This
is precisely twice the usual Sharvin conductance, and represents the 
current-doubling effect of Andreev reflection \cite{Beenakker}.

Eq. (\ref{bigdelI}) for the modification of the current due to the
impurity is still in a somewhat unwieldy form, due to the appearance
of the impurity lesser functions.  While the NS-NCA technique discussed
in this paper allows the calculation of spectral functions, it does
not allow one to calculate the distribution functions which are necessary
to compute the lesser Green function.
A similar problem occurs
when considering an impurity in a normal point contact \cite{hershfield},
or more generally, the current through an interacting region \cite{wingreen}.
In these cases, one can eliminate the lesser function by taking a suitable
linear combination of the current computed on the left of the interacting
region and that computed 
on the right.  Here, no such simplification is possible,
as computing the current within the superconductor is difficult due to 
the presence of supercurrents.  

A means for eliminating the lesser function does however exist.  We note
that the electron impurity Green functions $D_{11}(\omega)$ are related to the
hole impurity Green functions $D_{22}(\omega)$.  It follows easily from the
definitions of the functions that:

\begin{eqnarray}
\label{id1}
D^<_{22}(\omega) & = & -D^>_{11}(-\omega) \\
A_{22}(\omega) & = & A_{11}(-\omega)
\label{id2}
\end{eqnarray}

where the greater Green function $D^>_{11}$ is defined by:

\begin{equation}
D^>_{11}(t) = -i \left\langle d_{\uparrow}(t) d^\dag_{\uparrow}(0)
   \right\rangle
\end{equation}

We use Eqs. (\ref{id1}) and (\ref{id2}) to substitute for the third and
forth terms in Eq. (\ref{bigdelI}) for $\delta I$.  Further, we change
the integration variable for these terms from $\omega$ to $-\omega$, and
make the assumption that $\Gamma(\omega)$ is an even function.  The latter
is not too severe as long as the temperature and voltage are 
sufficiently small, as this ensures a small range of $\omega$ integration.
Finally, we use the relations $f_h(-\omega) = 1 - f_e(\omega)$ and
$D^>_{11} - D^<_{11} = -2\pi i A_{11}$ to obtain:

\begin{equation}
\label{delI}
\delta I = -\frac{4e}{h} \int d\omega (\pi \Gamma(\omega))
	A_d(\omega) \left( \frac{ f_e(\omega)-f_h(\omega) }{2} \right)
\end{equation}

Note that $A_d$ in the above formula is the same as $A_{11}$.  Taking $\Gamma$
to be a constant and differentiating with respect to voltage at V=0 
yields the formula (\ref{QPCCOND}) for the impurity contribution to the
point contact zero-bias conductance.  

We now comment on terms in the current which involve anomalous 
impurity green functions.  Retaining these terms and performing
manipulations similar to those described above leads to a second
correction term to the point contact current:

\begin{equation}
\delta I_{anom} = -\frac{4e}{h} \int d\omega (\pi \Gamma(\omega))
 (\frac{\omega}{\Delta} A_{12}(\omega)) 
 \left( \frac{ f_e(\omega)-f_h(\omega) }{2} \right)
\end{equation}

The change to the zero-bias conductance arising from this term is 0 at 0
temperature, and is generally much smaller than $\delta I$ in (\ref{delI})
for temperatures and voltages smaller than $\Delta$.
In the case of a strong-$U$ Anderson impurity, 
$A_{12}(\omega)$ is strongly suppressed due to the suppression of on-site
pair correlations.  Even in the absence of an on-site repulsion,
$A_{12}(\omega)$ is small for $\omega < \Delta$, as there are no 
non-evanescent quasiparticles in the gap.  The presence of the factor
$\frac{\omega}{\Delta}$ further reduces the contribution of this term for
small voltages and temperatures.  We thus neglect the contribution of this 
term in computing the point contact zero-bias conductance.  Note that we
have evaluated numerically the significance of this term in a few test cases
by calculating $A_{12}$ within the NS-NCA; it typically changes the
magnitude of the zero-bias anomaly by less than 0.5\%.

\end{multicols}


\begin{references}
\bibitem{largeN}
Though $1/N = 1/2$ is not a small parameter in the present context , 
large-$N$ approaches can still retain validity.  
As discussed in reference \cite{ncarefs},
they provide a systemaic means of organizing perturbation theory in a manner
which captures the essence of Kondo physics (i.e. the formation
of a singlet ground state in the 1-channel case).  This does not
occur if one attempts a perturbative expansion in the hopping parameter $W$.
\bibitem{BDG}
In the clean limit, the usual gapped density of states is only
recovered when one moves a coherence length into the superconductor, at
which point the contribution from evanescent quasiparticle modes is
negligible.
\bibitem{TKFORM}
Recall that for an Anderson impurity, $T_K$ is proportional to 
$\exp(-\pi |\varepsilon_d| / 2 \Gamma)$.
\end{references}
\end{document}